\documentstyle[11pt]{article}

\makeatletter

\def\dif{{\rm d}}
\def\deriv{\@ifnextchar[{\@deriv}{\@deriv[]}}
   \def\@deriv[#1]#2#3{\mathchoice%
{{\dif^{#1}#2\over\dif{#3}^{#1}}}{{\dif^{#1}#2/\dif{#3}^{#1}}}%
{{\dif^{#1}#2\over\dif{#3}^{#1}}}{{\dif^{#1}#2/\dif{#3}^{#1}}}}
\def\derpar#1#2{\mathchoice%
{{\partial#1\over\partial#2}}{{\partial#1/\partial#2}}%
{{\partial#1\over\partial#2}}{{\partial#1/\partial#2}}}
\def\dderpar#1#2#3{\mathchoice%
{{\partial^2 #1\over\partial #2\,\partial #3}}%
{{\partial^2 #1/\partial #2\,\partial #3}}%
{{\partial^2 #1\over\partial #2\,\partial #3}}%
{{\partial^2 #1/\partial #2\,\partial #3}}}
\def\secteqno{\@addtoreset{equation}{section}%
\def\theequation{\thesection.\arabic{equation}}}
\newcounter{subequation}
\def\thesubequation{\alph{subequation}}
\def\sneqnarray{\stepcounter{equation}\let\@currentlabel=\theequation
\setcounter{subequation}{1}
\def\@eqnnum{{\rm (\theequation.\thesubequation)}}
\global\@eqcnt\z@\tabskip\@centering\let\\=\@eqncr\let\@@eqncr=\@@sneqncr
$$\halign to \displaywidth\bgroup\@eqnsel\hskip\@centering
 $\displaystyle\tabskip\z@{##}$&\global\@eqcnt\@ne
 \hskip 2\arraycolsep \hfil${##}$\hfil
 &\global\@eqcnt\tw@ \hskip 2\arraycolsep $\displaystyle\tabskip\z@{##}$\hfil
  \tabskip\@centering&\llap{##}\tabskip\z@\cr}
\def\endsneqnarray{\@@sneqncr\egroup $$\global\@ignoretrue}
\def\@@sneqncr{\let\@tempa\relax
   \ifcase\@eqcnt \def\@tempa{& & &}\or \def\@tempa{& &}
   \else \def\@tempa{&}\fi
     \@tempa \if@eqnsw\@eqnnum\stepcounter{subequation}\fi
     \global\@eqnswtrue\global\@eqcnt\z@\cr}
\makeatother

\secteqno

\newcommand{\UBECM}
           {Departament d'Estructura i Constituents de la Mat\`eria\\
            Universitat de Barcelona\\
            Diagonal 647\\
            E-08028 Barcelona}
\newcommand{\CR}
           {Center for Relativity. Department of Physics\\
            University of Texas at Austin\\
             Austin, Texas 78712-1081}

\def\beq{\begin{equation}}
\def\eeq{\end{equation}}
\def\bea{\begin{eqnarray}}
\def\eea{\end{eqnarray}}
\def\beasn{\begin{sneqnarray}}
\def\eeasn{\end{sneqnarray}}
\def\beann{\begin{eqnarray*}}
\def\eeann{\end{eqnarray*}}

\begin{document}

\begin{center}
{\bf \Large Plugging the Gauge Fixing into the Lagrangian}

\vspace{36pt}

J M Pons

\vspace{24pt}

\UBECM

\vspace{12pt}

and

\vspace{12pt}

\CR

\vspace{36pt}
\end{center}
\thispagestyle{empty}

\begin{abstract}

A complete analysis of the consequences of introducing a set of
holonomic gauge fixing constraints (to fix the dynamics) into a
singular Lagrangian is performed.  It is shown in general that the
dynamical system originated from the reduced Lagrangian erases all the
information regarding the first class constraints of the original
theory, but retains its second class.  It is proved that even though the
reduced Lagrangian can be singular, it never possesses any gauge
freedom. As an application, the example of $n \cdot A = 0$ gauges in
electromagnetism is treated in full detail.

\end{abstract}

\vfill \hfill
\vbox{
}\null

\clearpage

\def\A{\alpha}

\section{Introduction}

Take the Lagrangian for electromagnetism, $ L = -\frac{1}{4} F_{\mu\nu}
F^{\mu\nu}$. $L$ is singular because its Hessian matrix with respect to
the velocities is singular.  Now consider a generic axial gauge
$n \cdot
A = 0$, with $n_0 \neq 0$, and plug this constraint into the Lagrangian
to eliminate $A^0$.  We end up with a reduced Lagrangian $L_R$ wich
turns out to be regular and that can be used to define the propagator.
The question is: is this procedure (plugging the constraint into the
Lagrangian) correct?.  Or put it in another way: what is the relation
between the dynamics defined by $L$ and that by $L_R$?.  Also: what
happened to the Gauss Law $\nabla \cdot {\bf E} = 0$, which is a
constraint for $L$ but is no longer present for $L_R$?.

In this paper we will give an answer to these questions by studying the
relations betwwen $L$ and $L_R$ in the general case of a Constrained
Dynamical System (CDS).  There is sometimes a bit of confusion when
using these words.  Here we mean dynamical systems defined through
a Lagrangian\footnote{The canonical
formalism is built out of it by using the Legendre transformation.},
that happens to be singular (i.e.: when the Hessian matrix of the
Lagrangian with respect to the velocities is singular or, equivalently,
when the Legendre transformation from velocity space to phase space is
not invertible). Constraints
naturally appear in the formalism as a consequence of the equations of
motion --if we are in velocity space-- or due to the fact that the
Legendre transformation is not invertible --if we are in phase space.
In phase space, the existence of constraints is compulsory, in velocity
space not.  The confusion can arise when one considers regular
(non-singular) Dynamical Systems which are deliberately constrained to
describe the motions in an {\it ad hoc} given surface.  This surface can
be either in configuration space (holonomic constraints) or in tangent
space or phase space.  In the holonomic case, one usually deals with
these systems with
the theory of Lagrange multipliers\cite{rund} \cite{sal1} \cite{sal2}
\cite{pepin}, which is physically based on D'Alembert's principle.  Let
us notice that in general the presence of these {\it ad hoc} constraints
ammounts to a change of the original equations of motion.

Dirac \cite{dirac50}\cite{dirac}, in his pioneering work, studied CDS in
order to get a Hamiltonian formulation of gauge theories,
including General Relativity (GR). As a
general covariant theory, GR contains some gauge transformations (i.e.:
symmetries that depend upon arbitrary functions of space-time.  In the
case of GR they are the spacetime diffeomorphisms) and it turns out that
the price for having this kind of transformations in the formalism is
that the Lagrangian must necessarily be singular.  In fact, not only GR,
but the most important quantum field theories, and also string theory,
have also room for gauge transformations.  This fact makes CDS a central
issue in the modern study of Dynamical Systems.

The existence of gauge transformations means that there are unphysical
degrees of freedom in the formalism.  This is also reflected in the fact
that there is some arbitrariness in the dynamics, since to a given set
of initial conditions there correspond several --actually, infinite--
solutions of the equations of motion, which are related among themselves
through gauge transformations.  To get rid of these transformations,
i.e., to quotient out the spurious gauge degrees of freedom, we must
somewhat reduce the dimensionality of tangent space or phase space
(depending upon we are working in Lagrangian or canonical formalism).
One way of doing that, which proves convenient in varied circumstances,
is by {\it ad hoc} introducing a new set constraints in order to
eliminate the unphysical degrees of freedom.  These constraints are
called {\it Gauge Fixing} (GF) constraints and its role is twofold
\cite{gra88} \cite{larry}: fixing the dynamics and setting the
physically inequivalent initial conditions.  Notice that now we are
introducing {\it ad hoc} constraints into the formalism as it can be
done for regular theories.  The difference is that now we do not intend
to modify the dynamics, but rather to fix it by selecting one specific
dynamics from the --gauge related-- family of possible dynamics
described by the equations of motion.

In this paper we will consider the GF procedure to fix the dynamics
\footnote{this is only a part of the whole GF procedure.  The second
part, as we have just said, consists in fixing the initial conditions.}
of a CDS in the case of an holonomic GF (i.e.: constraints defined in
configuration space $\cal Q$).  This is obviously the simplest case and
it allows for both Hamiltonian and Lagrangian analysis.  We find it
instructive to compare the role of these {\it ad hoc} GF constraints
introduced in a CDS with the role --dictated by D'Alembert principle--
of the {\it ad hoc} constraints introduced in a regular Dinamical
System. The
geometric version of D'Alembert principle is: The holonomic constraints
define a reduced configuration space $\cal Q_R$ which has a natural
injective map to $\cal Q$, $ i : \cal Q_R \longrightarrow \cal Q$. $i$
is naturally lifted to the tangent map between the tangent bundles, $ i'
:T \cal Q_R \longrightarrow T \cal Q$.  Then, the new dynamics is
defined in the velocity space $T \cal Q_R$ through a Lagrangian $L_R$
which is the image of $L$ (the original Lagrangian) under the pullback
$i'^*$ of $i'$.  In plain words this means that we get $L_R$ just by
substituting the constraints into $L$.

This is standard theory for constrained regular Dinamical Systems, of
course. But some
questions arise when we try to use the same mechanism in order to fix
the dynamics (through GF holonomic constraints) of a CDS.  Now, in the
singular case, we are lacking of any physical principle --like
D'Alembert's-- to justify this procedure, but since it is available, it
is worth to explore it.  The question is: Can we proceed in the same
lines as it is done in the regular case, i.e.: to produce a reduced
Lagrangian which dictates what the dynamics shall be?.  In other words:
it is correct to substitute the constraints into the lagrangian in order
to get the correct gauge fixed dynamics?  Is there any loss of
information of the original theory if we proceed this way?.  Our answer
will be that the procedure is correct but, since there really is a loss
of information, it must be supplemented with the addition of a specific
subset of the primary Lagrangian constraints of the original theory.

Throghout the paper we will work mainly in tangent space but we will
turn to the canonical formalism when we find it convenient.  Some
standard mathematical conditions for the Lagrangian are assumed, namely:
the rank of the Hessian matrix is constant and --in Hamiltonian
picture-- no second class constraint can become first class through the
stabilization algorithm.

In section 2 we will briefly consider the reduced Lagrangian formulation
of the system.  This formulation is equivalent to the extended one when
use is made of the Lagrange multipliers.  We will show in particular
that the Lagrange multipliers for a CDS GF constraints are combinations
of the original primary Lagrangian constraints of the theory.

The main result of this paper is (section 3) the following: Once the
holonomic GF constraints to fix the dynamics are introduced, a) The
reduced Lagrangian is singular if and only if the original theory has
Hamiltonian second class constraints, and b) there is no gauge freedom
for the reduced Lagrangian (i.e.: it has no first class Hamiltonian
constraints).

In section 4 we illustrate our results with some examples.  Comments and
conclusions are given in section 5.

\section{formalism in the reduced velocity space}

Let us first consider a time-independent Lagrangian $L(q,\dot q)$ in
$T{\cal Q}$, where $q = \{q^i, i=1,\cdots,n \}$ are local coordinates
for a point in a $n$-dimensional configuration space $\cal Q$.  At this
moment $L$ can be either regular or singular.

Let us introduce a set of --independent-- holonomic constraints
$f_{\mu}(q) = 0, \, \mu = 1,\cdots,k < n$.  These constraints define
a reduced configuration space $\cal Q_R$ where we can define coordinates
$ Q = \{Q^a,a=1,\cdots, N := n-k \}$.  The injective map $\cal Q_R
\rightarrow \cal Q$ is defined by some functions $q^i = q^i(Q^a)$ such
that the rank of $|\derpar{q^i}{Q^a}|$ is $k$ (maximum rank), and the
lifting of this map to the tangent structures allows to define the
reduced Lagrangian $L_R$ as
$$
L_R(Q,\dot Q) := L(q(Q), \derpar{q}{Q} \dot Q).
$$
Then, working in the second tangent bundles, it is easy to see that
\footnote{To alleviate the notation we sometimes do not
write the pullbacks explicitely. Here for instance the pullback of the
functions in $T^2\cal Q$ to functions in$T^2\cal Q_R$ is understood.}
$$
[L_R]_a = [L]_i \derpar{q^i}{Q^a},
$$
where $[L]_i$ stand for the functional derivatives of $L$:
$$
[L]_i := \derpar{L}{q^i} -
{d \over d \,t}
(\derpar{L}{\dot q^i} ) =:
\alpha_i - W_{ij} \ddot q^j.
$$

Since $\derpar{f_{\mu}}{q^i}$ form a basis for the independent null vectors of
$\derpar{q^i}{Q^a}$, i.e.,
\beq
\derpar{f_{\mu}}{q^i} \derpar{q^i}{Q^a} = 0
\label{ident}
\eeq
identically, we can conclude that there exists some
quantities $\lambda^{\mu}$ such that the equations $ [L_R]_a = 0$ are
equivalent to $ [L]_i + \lambda^{\mu} \derpar{f_{\mu}}{q^i} = 0 \, , \,
f_{\nu}(q) = 0 $, which in turn are equivalent to the equations obtained
from the variational principle for the extended lagrangian $L_E := L +
\lambda^{\mu} f_{\mu}$, where $\lambda^{\mu}$ are taken as new dynamical
variables.

So we have the well known result
$$
[L_R] = 0 \Longleftrightarrow [L_E] = 0,
$$
which is independent on whether $L$ is singular or not.  Let us first
consider the regular case.  The evolution operator in $T{\cal Q}$
derived from $L_E$ is
$$
{\bf X} = {\bf X_0}
+ \lambda^\mu\derpar{f_\mu}{q^i} (W^{-1})^{ij}\derpar{}{\dot q^j},
$$

where

$$
{\bf X_0} = \dot q^i \derpar{}{q^i} + \alpha_i
(W^{-1})^{ij}\derpar{}{\dot q^j},
$$

Stability of $\dot f_{\nu} := {\bf X}f_{\nu}= {\bf X_0}f_{\nu}$ under
${\bf X}$  determines the Lagrange multipliers as

$$
\lambda^{\nu} = - \theta^{\mu\nu} {\bf X_0}{\dot f_{\mu}},
$$

where $\theta^{\mu\nu}$ is the inverse of

$$
\theta_{\mu\nu} = \derpar{f_{\mu}}{q^i} (W^{-1})^{ij}\derpar{f_{\nu}}{q^j}.
$$

Observe that, except for the case when ${\dot f_{\mu}}$ is alreay
stable under ${\bf X_0}$, we end up with a dynamics which is different
to the original one.

Now for the singular case. The
equations of motion obtained from $L$ are:
$$	\alpha_i - W_{ij}\ddot q^j = 0 .
$$

Since $W_{ij}$ is singular, it possesses $r$ null vectors $\gamma^i_\rho$,
giving up to $r$ (independent or not) constraints
$$	\alpha_i \gamma^i_\rho = 0\ .
$$

It proves very convenient to use a basis for these null vectors which is
provided from the knowledge of the $r$ primary Hamiltonian constraints
of the theory, $\phi^1_\rho$. Actually one can take \cite{batlle86}:
\beq
\gamma^i_\rho= {\partial\phi^1_\rho\over\partial p_i}(q,\hat p)\ ,
\label{gamma}
\eeq
where $\hat p_i(q,\dot q)=\partial L/\partial\dot q^i$.
It is easily shown that there exists at least one $M^{ij}$ and
$\tilde\gamma^\rho_i$ such that
$$  \delta^j_i = W_{is}M^{sj}  + \tilde\gamma^\rho_i \gamma^j_\rho\ ,
$$
and therefore \cite{kam}
$$	\ddot q^i= M^{is}\alpha_s+ \tilde\eta^\rho\gamma^i_\rho\ ,
$$
where $\tilde\eta^\rho$ can be taken as arbitrary functions of $t$.

The stabilization algorithm starts by demanding that time evolution
preserve the constraints $\alpha_i\gamma^i_\rho$.
Sometimes new constraints are found;
sometimes some of the $\tilde\eta^\rho$ are determined; eventually the
dynamics is described by a vector field that exists on, and is tangent to,
the constraint surface in velocity space:
$${\bf X}:= {\partial\over\partial t}
		+ \dot q^i {\partial\over\partial q^i}
		+ a^i(q,\dot q){\partial\over\partial\dot q^i}
		+\eta^\mu\Gamma_\mu
	=: {\bf X}_0 + \eta^\mu\Gamma_\mu\ ;
$$
the $a^i$ are determined from the equations of motion and the
stabilization algorithm; $\eta^\mu$ ($\mu=1,\cdots,p_1$) are
arbitrary functions of time; and
$$	\Gamma_\mu = \gamma^i_\mu {\partial\over\partial\dot q^i}\ ,
$$
where $\gamma^i_\mu$ are a subset of the null vectors of
$W_{ij}$, corresponding to the  first class primary constraints $\phi_\mu$
\footnote{Here we refer to {\it first class primary constraints} as the subset
of primary constraints that still satisfy the first class condition {\it after}
we have run the stabilization algorithm to get all the constraints of the
theory: primary, secondary, etc. If we only look at the primary level, the
number of first class primary constraints can be greater.}
found in
the Hamiltonian formalism \cite{batlle86} (which we take in number $r_1
\leq r$).

A holonomic GF for the dynamics will consist in the introduction of
$r_1$ independent functions $f_{\mu}(q)$ in such a way that the arbitrary
functions $\eta^\mu$ ($\mu=1,\cdots,r_1$) of the dynamics become determined by
the requirement of stability of
$\dot f_{\nu} := {\bf X}f_{\nu}= {\bf X_0}f_{\nu}$ under the action of
${\bf X}$. To do so we need:
\beq
|\Gamma_\mu \dot f_{\nu}| = |\gamma^i_{\mu} \derpar{f_{\nu}}{ q^i} |
= |\{\phi^1_{\mu}, f_\nu  \}| =: |D_{\mu\nu}| \neq 0.
\label{det}
\eeq
(observe that the relation
$|\{f_\nu ,\phi^1_{\mu} \}|  \neq 0$ shows
that the GF constraints $f_\mu = 0$ also fix the dynamics in canonical
formalism, since
$\phi^1_\mu$ are the first class primary constraints in the Hamiltonian
formalism.).

With these GF constraints $f_{\mu}(q)$, the extended Lagrangian
$L_E$ gives the equations of motion:

$$
[L]_i + \lambda^{\mu} \derpar{f_{\mu}}{q^i} = 0 \, , \, f_{\nu}(q) = 0.
$$

Now compute $\lambda^{\mu}$.
Contraction of $\gamma^i_\nu$ with the first set of these equations
gives
$$
\alpha_i \gamma^i_\nu + \lambda^\mu \gamma^i_\nu\derpar{f_{\mu}}{q^i}
= \alpha \gamma_\nu + \lambda^\mu D_{\nu\mu}\ ,
$$
(we have supressed some coordinate indices in the last expression)
whereby we can get $\lambda^\mu$ as
\beq
\lambda^\mu = - (D^{-1})^{\mu\nu} \alpha \gamma_\nu.
\label{lambda}
\eeq

The noticeable fact is that now the Legendre multipliers $\lambda^\mu$ are
constraints. This is good news because it tells us that the introducction of
the GF constraints has not modified the dynamics. Let us be more specific on
this point.

There is a splitting of the set of hamiltonian primary constraints as first
and second class (we will use indices $\mu$ for first class, and $\mu'$ for
second class) constraints\footnote{The previous footnote applies here.}.
It will prove convenient later, since the first class
constraints $\phi^1_\mu$ satisfy  $|\{f_\nu ,\phi^1_{\mu} \}| \neq 0$, to
take a basis $\phi^1_{\mu'}$ for the second class constraints such that
\beq
\{\phi^1_{\mu'}, f_\nu  \} = 0\ ;
\label{phipro}
\eeq
then
\beq
\gamma_{\mu'}^i \derpar{f_{\nu}}{q^i}  = 0\ .
\label{fg}
\eeq
This splitting ($\mu,\, \mu'$) is
translated to the Lagrangian formalism through (\ref{gamma}), so we end
up with primary Lagrangian constraints $\alpha \gamma_\mu \simeq 0$ and
$\alpha \gamma_{\mu'} \simeq 0$. It is worth noticing that only the
first set is involved in the determination of $\lambda^\mu$.

Now observe that, due to (\ref{lambda}):
$$
[L] + \lambda^{\mu} \derpar{f_{\mu}}{q} = 0 \, , \,
\alpha \gamma_\mu = 0
\Longleftrightarrow
[L] =0 \, , \, \alpha \gamma_\mu = 0\ ,
\Longleftrightarrow
[L] =0
$$
the last equality holding because the constraints $\alpha \gamma_\mu =0$
are consequence
of $[L] = 0$. Therefore
$$
[L] = 0 \, , \, f_{\mu}(q) = 0
\Longleftrightarrow
$$
$$
[L] + \lambda^{\mu} \derpar{f_{\mu}}{q} = 0 \,  ,  \,
\alpha \gamma_\mu = 0 \, , \,f_{\mu}(q) = 0
 $$
\beq
\Longleftrightarrow
[L_E] = 0 \, , \, \alpha \gamma_\mu = 0
\Longleftrightarrow
[L_R] = 0 \, , \, \alpha \gamma_\mu = 0\ ,
\label{llr}
\eeq
where $\alpha \gamma_\mu$ is obviously understood with the pullback
to $T \cal Q_R$, $i'^* (\alpha \gamma_\mu)$.

So we see that $[L] = 0 \, , \, f_{\mu}(q) = 0
\Longrightarrow [L_R] = 0$ but the converse is
not true. In the next section we will get a perfect understanding of
this fact.

\section{When is $L_R$ regular?}

With the same notation as in the previous section, with $L$ being a singular
Lagrangian and $L_R$ the reduced Lagrangian after an holonomic gauge fixing,
we are going to prove the following
\proclaim Theorem 1.
{\it$L_R$ is regular if and only if $L$ has only first class (Hamiltonian)
constraints.}

\subsection{Proof of theorem 1}

 First observe that we can write
\beq
\dderpar{L_R}{\dot Q^a}{\dot Q^b} = W_{ij} \derpar{q^i}{Q^a} \derpar{q^j}{Q^b}\
,
\label{hess}
\eeq
where $W_{ij}$ stands, as before, for $\dderpar{L}{\dot q^i}{\dot q^j}$.
We must check whether the Hessian matrix for $L_R$ is regular or not.
That is, look
for the existence of solutions $V^a$ of
$(\dderpar{L_R}{\dot Q^a}{\dot Q^b}) V^b = 0$.
Then, using (\ref{hess}):
$$
0 = V^a \derpar{q^i}{Q^a} \derpar{q^j}{Q^b} W_{ij} =
\derpar{q^j}{Q^b}(W_{ij} \derpar{q^i}{Q^a} V^a)\ ,
$$
and since $\derpar{f_{\mu}}{q^j}$ form a basis for the null vectors of
$\derpar{q^j}{Q^b}$, there must exists $\eta^\mu$ such that
$$
W_{ij}\derpar{q^i}{Q^a} V^a = \eta^\mu \derpar{f_{\mu}}{q^j}\ ;
$$
contraction with $\gamma^j_\nu$, which are null vectors for $W_{ij}$, gives:
$$
0 = \eta^\mu \gamma^j_\nu\derpar{f_{\mu}}{q^j} = \eta^\mu D_{\nu\mu}\ ,
$$
but since $|D_{\nu\mu}| \neq 0$ we conclude that $\eta^\mu = 0$, which means
\beq
W_{ij}\derpar{q^i}{Q^a} V^a = 0\ .
\label{w}
\eeq
 Now, since the set of null vectors of
$W_{ij}$ is $\gamma^j_\nu \, , \, \gamma^j_{\nu'} \, $, from (\ref{w}):
$$
\derpar{q^i}{Q^a} V^a =
\delta^\nu \gamma^i_\nu + \delta^{\nu'} \gamma^i_{\nu'}\ ,
$$
for some $\delta^\nu \, , \, \delta^{\nu'}$.
 Contraction of this last expression
with $\derpar{f_{\mu}}{q^i}$, and use of (\ref{ident}) and
 (\ref{fg}), gives:
$$
\delta^\nu \gamma^i_\nu \derpar{f_{\mu}}{q^i}  = \delta^\nu D_{\nu\mu} = 0
\Longrightarrow  \delta^\nu = 0\ ,
$$
where (\ref{det}) has also been used.
Therefore
$$
\derpar{q^i}{Q^a} V^a = \delta^{\nu'} \gamma^i_{\nu'}\ ,
$$
for $\delta^{\nu'}$ arbitrary. This means that we will get as many independent
null vectors $V$ for $\dderpar{L_R}{\dot Q^a}{\dot Q^b}$ as indices run for
$\nu'$, which is the number of second class primary Hamiltonian constraints.
This
proves the theorem, for if there are no second class primary constraints, there
are no second class constraints at all. In such a case, there are no null
vectors
for $\dderpar{L_R}{\dot Q^a}{\dot Q^b}$ and $L_R$ is regular.

To obtain a basis for the null vectors $V^a$ is is convenient to have a deeper
look at the canonical formalism.

\subsection{canonical formalism}

Here we will only consider some results that prove interesting to us. There
is a natural map
$$ \tilde i  :T^*_{q(Q)} \cal Q  \longrightarrow  T^*_Q \cal Q_R.
$$
Momenta
$p_i$ are mapped to momenta $P_a$ of the reduced formalism according to
$P_a = p_i \derpar{q^i}{Q^a}$. Functions
$\psi(q,p)$ in $T^* \cal Q$ are projectable to functions $\tilde \psi(Q,P)$
in $T^* \cal Q_R$ if and only if
\beq
\psi(q(Q),p) =\tilde \psi(Q,p \derpar{q}{Q})
\label{tilde}
\eeq
This projectability condition can be writen in a more familiar way.
{}From (\ref{tilde}):
$$
\derpar{\psi}{p_i} = \derpar{\tilde \psi}{P_a} \derpar{q^i}{Q^a}.
$$
Contraction with
$\derpar{f_\mu}{q^i}$, and use of (\ref{ident}) gives
$$
\derpar{\psi}{p_i} \derpar{f_{\mu}}{q^i} = \{f_\mu ,\psi \}|_
{(f_\mu = 0)} = 0.
$$
This is the version of the projectability condition for $\psi$
we were looking for:
\beq
 \exists \, \tilde\psi, \, \tilde i^* : \psi \longrightarrow \tilde\psi
\Longleftrightarrow \{f_\mu ,\psi \}|_{(f_\mu = 0)} = 0
\label{project}
\eeq
So, according to (\ref{phipro}), the second class primary constraints
$\phi^1_{\mu'}$ are projectable to some functions  $\tilde \phi^1_{\mu'}$.
We will prove that $\tilde \phi^1_{\mu'}$ are the primary Hamiltonian
constraints coming from the reduced Lagrangian $L_R$. First we can see that
they are in the right number because it coincides with the number of null
vectors $V$ we have found for the Hessian matrix of $L_R$. Thererfore we only
have to check that the pullback of $\tilde \phi^1_{\mu'}$ to the tangent
space $T \cal Q_R$ is identically zero (this is the definition of primary
constraints). In fact:
$$
\tilde \phi^1_{\mu'}(Q, \derpar{L_R}{\dot Q}) =
\tilde \phi^1_{\mu'}(Q, \derpar{L}{\dot q} \derpar{q}{Q}) =
\phi^1_{\mu'}(q(Q), \derpar{L}{\dot q}) = 0
$$
identically.
We have thus got a basis for the null vectors $V$ as
$V_{\mu'}^a = \derpar{\tilde \phi^1_{\mu'}}{P_a} := \tilde \gamma_{\mu'}^a$.

Now we can get the primary Lagrangian constraints for $L_R$.
It is easy to see that
$$
{\alpha_R}_a := \derpar{L_R}{Q^a} - \dot Q^b \dderpar{L_R}{Q^b}{{\dot Q}^a}
= (\alpha_i + W_{ij} \dot Q^c \dderpar{q^j}{Q^b}{Q^c}) \derpar{q^i}{Q^a}\ ;
$$
then the primary Lagrangian constraints $\alpha_R\tilde \gamma_{\mu'}$
for $L_R$ are
$$
{\alpha_R}_a \tilde \gamma_{\mu'}^a =
(\alpha_i + W_{ij} \dot Q^c \dderpar{q^j}{Q^b}{Q^c}) \derpar{q^i}{Q^a}
\derpar{\tilde \phi_\mu'}{P_a} =
$$
$$
(\alpha_i + W_{ij} \dot Q^c \dderpar{q^j}{Q^b}{Q^c})
\derpar{ \phi_\mu'}{p_i} =  \alpha_i \derpar{ \phi_\mu'}{p_i}
= \alpha\gamma_\mu'\ ,
$$
where obvious pullbacks to $T \cal Q_R$ are understood.

We now have a complete understanding of (\ref{llr}):
\beq
[L] = 0 \, , \, f_{\mu}(q) = 0
\Longleftrightarrow
[L_R] = 0 \, , \, \alpha \gamma_\mu = 0 \, ,
\label{equiv}
\eeq
for the dynamics for $L_R$ only provides with the primary constraints
$\alpha_R\tilde \gamma_{\mu'}$, whereas the rest of primary constraints for
$L$,
$\alpha \gamma_\mu$, are absent in the reduced formulation, and must be
separately introduced in order to maintain equivalence with
$[L] = 0 \, , \, f_{\mu}(q) = 0$.

Now we are ready to prove:

\proclaim Theorem 2.
{\it The dynamics derived from $L_R$ has no gauge freedom.}

This is equivalent to say that all Hamiltonian constraints for $L_R$ are
second class.

\subsection{Proof of theorem 2}

First we will get the canonical Hamiltonian for $L_R$.
The dynamics determined by the GF $f_\mu = 0$ is described in the canonical
formalism by a first class Hamiltonian $H_{FC}$ which can be taken
to satisfy $\{ f_\mu,H_{FC}\} = 0$ (Since the stabilization of $f_\mu = 0$
determines $H_{FC}$). According to (\ref{project}), $H_{FC}$ is
projectable to a function $H_R := \tilde H_{FC}$ in $T^* \cal Q_R$

We can prove that
$H_R$ is a
canonical Hamiltonian corresponding to $L_R$. First consider the fact,
very easy to verify, that
the Lagrangian energy $E_R :=  \dot Q \derpar{L_R}{\dot Q} - L_R$ satisfies
$E_R := i'^*(E)$, where $E$ is tha Lagrangian energy for $L$. Now, defining
$FL: T \cal Q \longrightarrow T^* \cal Q$ as the Legendre
map derived from $L$, and $FL_R: T \cal Q_R \longrightarrow T^* \cal Q_R$ as
the Legendre map from $L_R$,
then the following property holds: for any function $\psi$ in $T^* \cal Q$
projectable to $\tilde \psi$ in $T^* \cal Q_R$,
$FL_R^* \tilde \psi = i'^* FL^* \psi$.
Then $E_R = i'^*(E) = i'^* FL^* H_{FC} = FL_R^* \tilde H_{FC}$, which proves
that $H_R := \tilde H_{FC}$ is a good Hamiltonian for $L_R$.

Next, it is easy to see that for two projectable functions, $\psi \, , \xi$,
(see (\ref{project})), its Poisson Bracket (PB) is projectable and
satisfies: \beq
\widetilde{ \{\psi,\xi \} } = \{\tilde \psi,\tilde \xi \}_R\ ,
\label{pb}
\eeq
where $\{\, , \}_R$ stands for the PB in $T^* \cal Q_R$
(The proof of (\ref{pb}) is immediate).

Now we can realize that the stabilization algorithm for the second class
primary constraints $\phi_{\mu'}$ in $T^* \cal Q$ is exactly the same as for
the primary constraints $\tilde \phi_{\mu'}$ in $T^* \cal Q_R$. Indeed, the
time derivative of $ \tilde \phi^1_{\mu'}$ is:
$$
\dot {\tilde \phi^1}_{\mu'} = \{\tilde \phi^1_{\mu'},H_R\}_R =
\widetilde{ \{ \phi^1_{\mu'},H_{FC}\} }=: \tilde \phi^2_{\mu'}\ ,
$$
and so on.
But since the set of constraints $\phi^1_{\mu'}, \phi^2_{\mu'}, \cdots$
(until the stabilization algorithm eventualy ends when no new constraints
appear) is second class, we conclude, using (\ref{pb}) that so it must
be for
$\tilde \phi^1_{\mu'}, \tilde \phi^2_{\mu'}, \cdots$. Hence Theorem 2 has been
proved. Notice that $H_R$ actually is a first class Hamiltonian.

The final picture for the reduction $L \longrightarrow L_R$ through the
GF constraints $f_\mu = 0$ has been established: In the reduced theory
there
is only place for the second class Hamiltonian constraints of the original
theory. First
class constraints have simply disappeared. Not only the primary ones
$\psi^1_\mu$ but also the secondary $\psi^2_\mu$, etc. that arise through
the application of the stabilization algorithm to $\psi^1_\mu$. Consequently,
the pullbacks of these first class constraints to velocity space, also
disappear from the reduced velocity space. All the
information carried by the first class structure has been erased. This applies
in particular to
the gauge generators, which are special combinations
--with arbitrary functions and its time derivatives as coeficients-- of
the first class constraints.

\section{Examples}
\subsection{Axial gauges in electromagnetism}
In the introduction  we have mentioned the example of electromagnetism in
$n \cdot A = 0$ gauges. $n^\mu = (n^0,{\bf n}), A^\mu = (A^0,{\bf A})$.
After elimination of $A^0$ the Lagrangian becomes
regular. Gauss law is missing. But there is a compatibility between this
missing constraint and the reduced dynamics. Now the time evolution vector
field is tangent to the Gauss law, and if our initial (for, say, time t=0)
configuration of the field ${\bf A}$ satisfies $\nabla {\bf E} = 0$, then the
constraint will be satisfied for all times. In QED we can use the reduced
Lagrangian to write down the propagator for the photon in the gauge
$n \cdot A = 0$, but to settle the asymptotic initial and final states, we
must require the fulfillment of Gauss law \cite{vannen}.

Let $L = - \frac{1}{4} F_{\mu\nu}F^{\mu\nu}$ be our starting point
(Our metric is (+,-,-,-)).
The only Lagrangian constraint is $ \alpha \gamma =
\partial_0(\partial_i A^i) + \Delta A_0  = 0$, the Gauss law.
The time evolution operator, which only exists on, and is tangent to,
the surface defined by the Gauss law, is:
$$
{\bf X} = \int d^3 y \, \dot A^\mu {\delta \over \delta A^\mu({\bf y})}
- \int d^3 y \,(\partial_j(\partial_\mu A^\mu) + \Delta A_j)
{\delta \over \delta \dot A^j({\bf y})} +
\int d^3 y \, \lambda({\bf y},t) {\delta \over \delta \dot A^0({\bf y})},
$$
where $\lambda$ is the arbitrary function of the dynamics reflecting the
existence of gauge freedom. The GF $n \cdot A=0$ will fix $\lambda$ as
$$
\lambda = -\frac{1}{n^0}( {\bf n \cdot \nabla}(\partial_\mu A^\mu)
+ \Delta ({\bf n \cdot A})).
$$
If the variable $A_0$ is eliminated by using the GF constraint, we end up with
the reduced evolution operator --always tangent to the Gauss law constraint--
\beq
{ \bf \hat X} = \int d^3 y \, \dot A^j {\delta \over \delta A^j({\bf y})}
- \int d^3 y (\partial_j(\partial_i A^i + \frac{1}{n^0}
\partial_0({\bf n \cdot A})) + \Delta A_j)
{\delta \over \delta \dot A^j({\bf y})}.
\label{em}
\eeq

Now let us proceed the other way around, that is, instead of substituting the
GF constraint $n \cdot A = 0$ into the equations on motion, let us plug
$n \cdot A = 0$ into the Lagrangian $L$ to get the reduced  Lagrangian
$L_R[{\bf A}, {\bf \dot A}]$:
$$
L_R =- \frac{1}{4} F_{ij}F^{ij} -  \frac{1}{2} F_{0i}F^{0i},
$$
with $F_{0i} = \dot A_i - {1 \over n^0} \partial_i ({\bf n \cdot A}).$
This Lagrangian is regular and its time evolution operator is:
\pagebreak
$$
{\bf X_R} = \int d^3 y \, \dot {\bf A}^j {\delta \over \delta {\bf A}^j({\bf
y})}
$$
\beq
- \int d^3 y \,
(\Delta A_j +
\frac{1}{n^0} (\partial_0 \partial_j + \frac{n^j}{n^0}\Delta)
({\bf n \cdot A}) \, +
(\partial_j +  \frac{n^j}{n^0} \partial_0) (\partial_i A^i))
{\delta \over \delta \dot {\bf A}^j({\bf y})}
\label{em2}
\eeq

As it was expected, there is no trace of the Gauss law.
We can check, though, that the evolutionary vector field ${\bf X_R}$ in
(\ref{em2})
differs from the evolutionary vector field ${\bf {\hat X}}$ in
(\ref{em}) by a term which is
proportional to the Gauss law. Indeed:

$$
{ \bf \hat X} - {\bf X_R} =
\int d^3 y \,
\frac{n^j}{n^0} (\partial_0(\partial_i A^i) + \frac{1}{n^0} \Delta (
{\bf n \cdot A}))
{\delta \over \delta \dot {\bf A}^j({\bf y})}
$$
This result explicitly exemplifies the relation (\ref{equiv}).

\subsection{Pure Abelian Chern-Simons}

Consider the Abelian Chern-Simons 2+1 Lagrangian
$L = \frac{1}{2} \epsilon ^{\mu\nu\rho} F_{\mu\nu} A_\rho$ (no metric
involved). The primary hamiltonian constraints are
$$
\phi^\sigma := \pi^\sigma - \epsilon^{0\sigma\rho}A_\rho = 0.
$$
$\phi^0$ is a first class constraint, whereas $\phi^1, \, \phi^2$ are second
class. There is a secondary first class constraint
$\psi := \partial_i (\pi^i + \epsilon^{0ij}A_j) = 0$ which plays the role of
the Gauss law in this case.

Use of the GF constraint $A^0 = 0$ to get the reduced Lagrangian $L_R$ gives:
$$
L_R = \epsilon^{0ij} \dot A_i A_j,
$$
which is still singular. Its only primary constraints are second class:
$\tilde \phi^1 := \pi^1 - A_2 = 0, \,  \tilde \phi^2 := \pi^2 + A_1 = 0$.
No secondary constraints arise and Gauss law has disappeared.
Observe that we are verifying our {\bf theorem 2}: the reduced Lagrangian
$L_R$ is still singular (because $L$ has second class constraints) but has
no room for gauge freedom.

\section{conclusions}

In this paper we have proved some results concerning the correctness of
plugging a set of holonomic GF constraints into a singular Lagrangian
$L$.  Our result is: if the GF constraints are taken in such a way that
properly fix the dynamics of the singular theory defined by $L$, then
the reduced Lagrangian $L_R$ only keeps the information of the second
class Hamiltonian constraints of the original theory.  All first class
constraints have disappeared from the reduced formalism.  Therefore, to
maintain equivalence with the dynamics defined by $L$ we must add the
primary first class constraints to the dynamics defined by $L_R$.  It is
remarkable that all the information carried by the first class structure
has been erased.  No only at the primary level, but at any level of the
stabilization algorithm.  This applies in particular to the gauge
generators, which are made up of the first class constraints.

This is the picture in Hamiltonian formalism.  In Lagrangian formalism
the constraints erased through this procedure of plugging the GF
constraints into the Lagrangian are just the pullbacks to velocity space
of the first class Hamiltonian constraints of the original theory.

Lorentz covariance of Yang-Mills type theories requires the existence of
secondary Hamiltonian constraints (whose pullbacks to velocity space
will be part of the primary Lagrangian constraints).  Also general
covariance for theories containing more that scalar fields (like General
Relativity) require these secondary constraints also.  Therefore, in all
these cases, the price for plugging the holonomic GF constraints into
the Lagrangian will be the loss of the secondary first class
constraints.  The examples provided in the previous section not only
show how this happens but also show the consistency of the reduced
theory with the missing constraints.

In our paper we have dealt with holonomic GF constraints only.  The
advantage of considering this case is that we have to our avail a
reduced configuration space out of which both the reduced velocity and
phase space are built.

Let us finish with two more comments.  First, the theory we have
developped can be easily extended to cases where only a partial gauge
fixing of the dynamics is performed.  In such cases, there is still some
gauge freedom left for $L_R$ and only a part of the original primary
first class constraints --and its descendants through the stabilization
algorithm-- disappear from the reduced formalism.  Second, In the usual
case (Yang-Mills, Einstein-Hilbert gravity, etc.) where the
stabilization algorithm in velocity space has only one step,
the surface defined by $\alpha \gamma_\mu = 0$
(see Eq. (\ref{equiv})) is a constant of motion for $L_R$, but
not
a standard one, since $\alpha \gamma_\mu = c \neq 0$) will not be, in
general, a constant of motion.  Our first example can be used to
illustrate this aspect.

\section{Acknowledgements}

This work has been partially supported by the CICIT (project number AEN-0695)
and by a Human Capital and Mobility Grant (ERB4050PL930544). The author thanks
the Center for Relativity at The University of Texas at Austin for
its hospitality.

\end{document}